\documentclass[sigconf]{acmart}

\usepackage{graphicx}
\usepackage{amsmath}
\usepackage{booktabs}
\usepackage{multirow}
\usepackage{algorithm}
\usepackage{algorithmic}
\usepackage{caption}

\AtBeginDocument{%
  \providecommand\BibTeX{{%
    \normalfont B\kern-0.5em{\scshape i\kern-0.25em b}\kern-0.8em\TeX}}}

\begin{document}

\title{Go Wide or Go Deep: Levering Watermarking Performance with Computational Cost for Specific Images}


\author{Zhaoyang Jia}
\email{jzy_ustc@mail.ustc.edu.cn}
\affiliation{%
  \institution{University of Science and Technology of China}
  \streetaddress{No. 443, Huangshan Road}
  \city{Hefei}
  \state{Anhui}
  \country{China}
  \postcode{230027}
}

\author{Han Fang}
\authornote{Corresponding author.}
\email{fanghan@mail.ustc.edu.cn}
\affiliation{%
  \institution{University of Science and Technology of China}
  \streetaddress{No. 443, Huangshan Road}
  \city{Hefei}
  \state{Anhui}
  \country{China}
  \postcode{230027}
}

\author{Zehua Ma}
\email{mzh045@mail.ustc.edu.cn}
\affiliation{%
  \institution{University of Science and Technology of China}
  \streetaddress{No. 443, Huangshan Road}
  \city{Hefei}
  \state{Anhui}
  \country{China}
  \postcode{230027}
}

\author{Weiming Zhang}
\authornotemark[1]
\email{zhangwm@ustc.edu.cn}
\affiliation{%
  \institution{University of Science and Technology of China}
  \streetaddress{No. 443, Huangshan Road}
  \city{Hefei}
  \state{Anhui}
  \country{China}
  \postcode{230027}
}



\begin{abstract}
    Digital watermarking has been widely studied for the protection of intellectual property. Traditional watermarking schemes often design in a "\textit{wider}" rule, which applies one general embedding mechanism to all images. But this will limit the scheme into a \textit{robustness-invisibility} trade-off, where the improvements of robustness can only be achieved by the increase of embedding intensity thus causing the visual quality decay.  However, a new scenario comes out at this stage that many businesses wish to give high level protection to specific valuable images, which requires high robustness and high visual quality at the same time. Such scenario makes the watermarking schemes should be designed in a "\textit{deeper}" way which makes the embedding mechanism customized to specific images. To achieve so, we break the robustness-invisibility trade-off by introducing computation cost in, and propose a novel auto-decoder-like image-specified watermarking framework (ISMark). Based on ISMark, the strong robustness and high visual quality for specific images can be both achieved. In detail, we apply an optimization procedure  (OPT) to replace the traditional embedding mechanism. Unlike existing schemes that embed watermarks using a learned encoder, OPT regards the cover image as the optimizable parameters to minimize the extraction error of the decoder, thus the features of each specified image can be effectively exploited to achieve superior performance. Extensive experiments indicate that ISMark outperforms the state-of-the-art methods by a large margin, which improves the average bit error rate by 4.64\% (from 4.86\% to 0.22\%) and PSNR by 2.20dB (from 32.50dB to 34.70dB).
\end{abstract}

\settopmatter{printacmref=false} 
\renewcommand\footnotetextcopyrightpermission[1]{} 
\pagestyle{plain} 


\maketitle

\section{Introduction}
\label{sec:intro}

\subsection{Digital Watermarking : Go Wide or Go deep?}
\label{sec:1-1}

\begin{figure}[t]
    \centering
    \includegraphics[width=\linewidth]{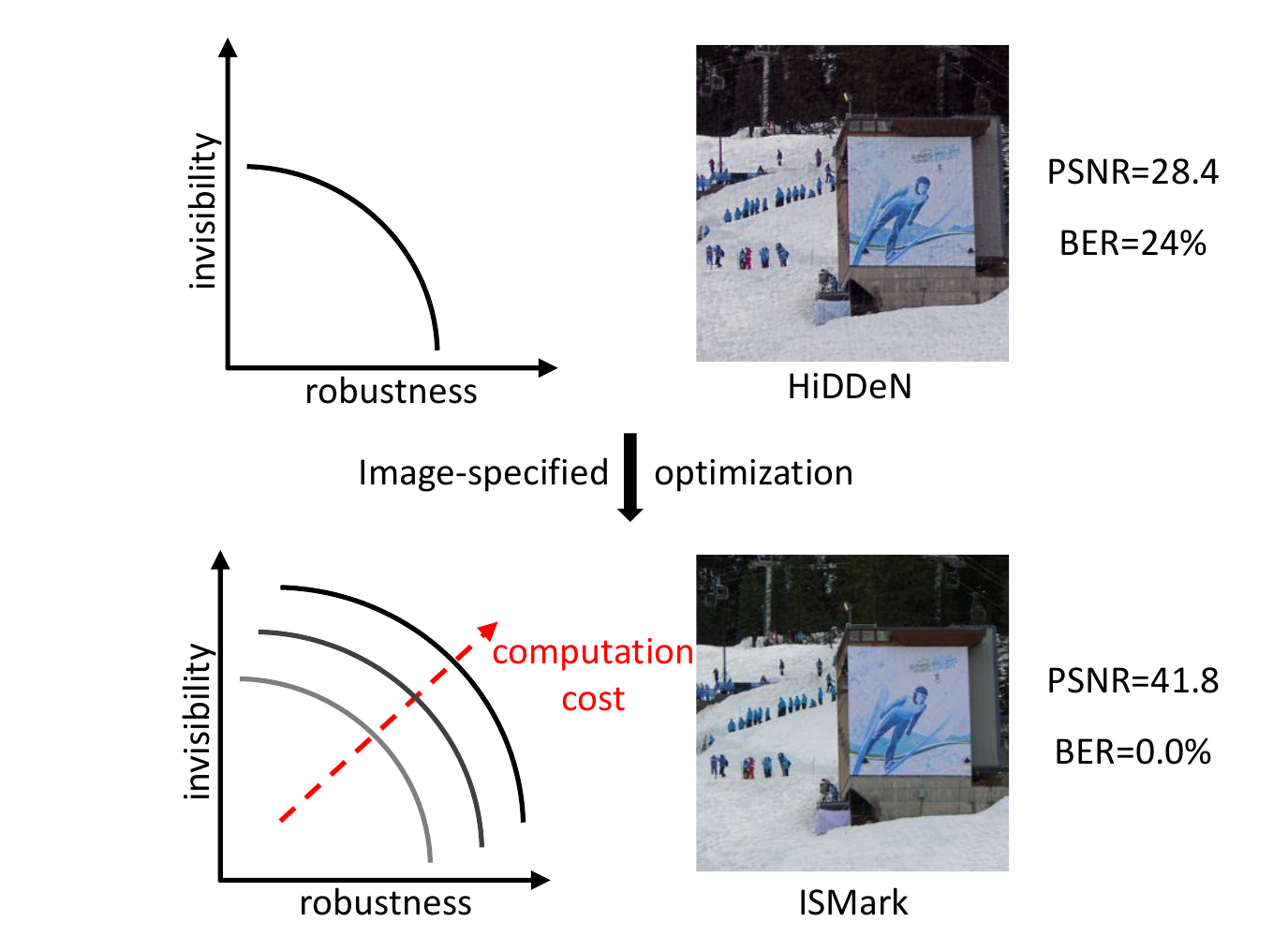}
    
    \caption{We break the robustness-invisibility trade-off by introducing computational cost in, and the proposed image-specified watermarking framework (ISMark) can greatly lever watermarking performance with computational cost. In the right we show an example, where both the visual quality (PSNR) and the robustness (bit error rate, BER) are enhanced with the image-specified optimization. Best view the details of the images in zoom in.}
    \label{fig1}
\end{figure}

Digital watermarking is an important technology for intellectual property protection. By embedding the identification information (i.e.,  a \textit{watermark}) into the \textit{cover image}, the ownership of the image can be confirmed. In this process, there are two basic requirements: 1) \textbf{invisibility}, which requires that no serious visual distortion occurs during watermark embedding so that the watermark will not be simply observed, and 2) \textbf{robustness}, which ensures the accuracy of watermark extracting even if the image is severely distorted. 

Recently, a growing number of schemes have been proposed to embed watermarks using either handcrafted modules\cite{van1994digital, tirkel1995two, bao2005image, karybali2006efficient, nasir2007new,bi2007robust, makbol2013robust, wang1998wavelet, ma2021local} or learning-based neural networks\cite{zhu2018hidden, liu2019novel, jia2021mbrs,fang2022encoded} and have achieved impressive results. These schemes aim to "\textbf{go wide}" to design an universal rule for watermarking. That is, they study on a set of images or possible attacks to design one general embedding mechanism for all images. But it will limit the designed watermarking scheme into a robustness-invisibility trade-off, since under these general embedding mechanisms, the improvement of robustness can only be achieved by the increase of embedding intensity, which causes the reduction of visual quality.

However, a new scenario comes out at this stage that many businesses wish to give high level protection to specific valuable images  (e.g. carefully designed promotion pictures),  which requires high robustness and high visual quality at the same time. Apparently, existing watermarking scheme cannot be well applied to such scenario due to the limitation of the robustness-invisibility trade-off. 

To meet the needs of such scenarios, the watermarking schemes should "\textbf{go deep}" to be customized to specific images. It means we should exploit image-specified features to design watermarks according to characteristic of each image to achieve better performance. To achieve so, we propose a novel image-specified watermarking framework (ISMark), which introduce the \textit{computation cost} in as a new metric to break the \textit{robustness-invisibility} trade-off of digital watermarking. As shown in Fig.\ref{fig1}, with the proposed ISMark, both high invisibility and robustness can be achieved.

\subsection{Image-Specified Watermarking Framework}
\label{sec:1-2}

Unlike previous schemes that use a general embedding mechanism to all images, ISMark exploits image-specified features to design watermarks for each image. In detail, we apply an optimization procedure (OPT) in embedding, which regards the cover image as optimizable parameters to minimize the extraction error of the decoder. With OPT embedding, we can enhance the watermarking performance by increasing the optimization steps, and as a result the robustness-invisibility can be both achieved with the increment of computation cost.

\begin{figure}[t]
    \centering
    \includegraphics[width=\linewidth]{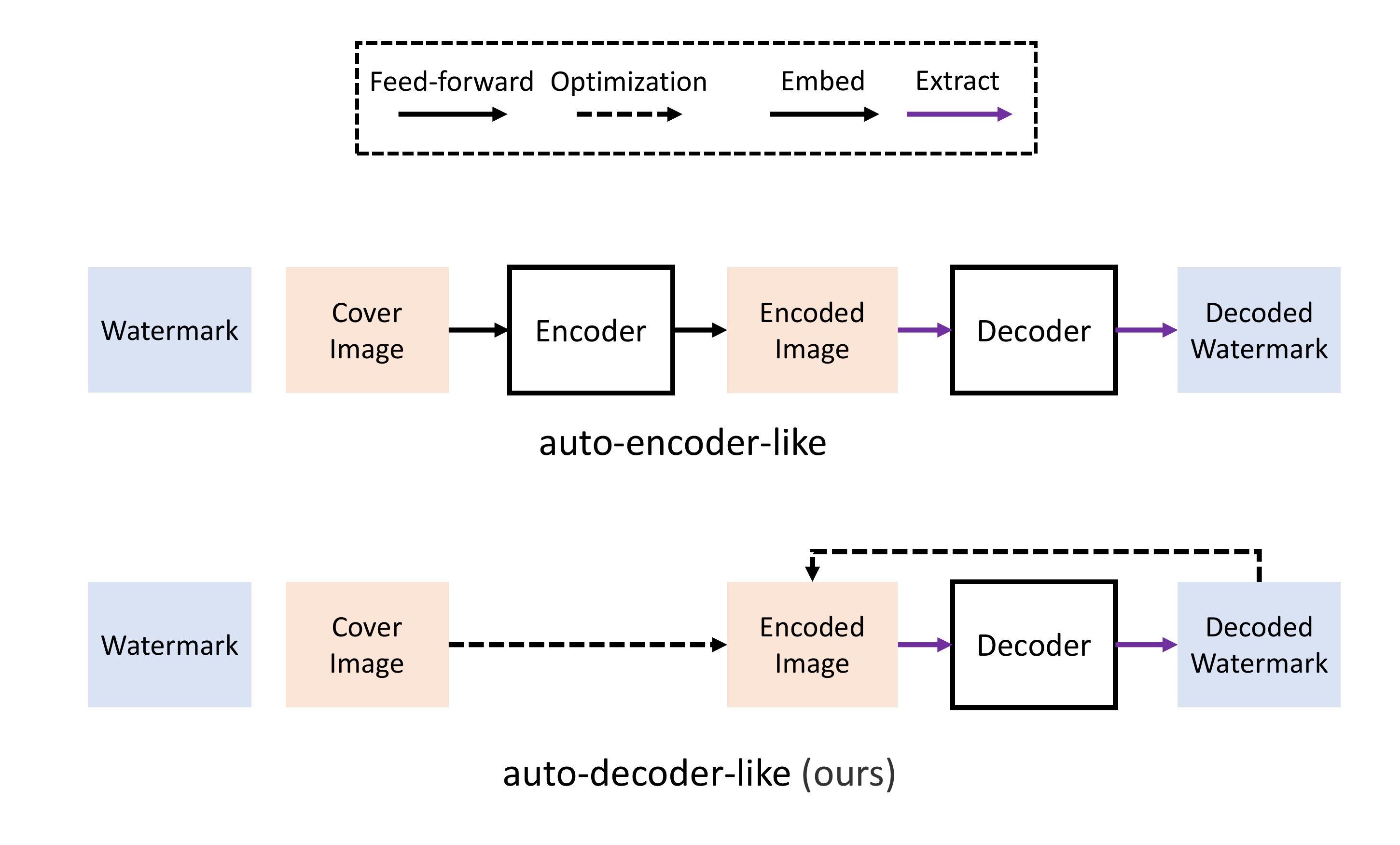}
    
    \caption{Framework structure comparison. Previous methods adopt an auto-encoder-like structure and embed watermarks using an encoder, while our auto-decoder-like structure embed watermarks through optimization.}
    \label{fig2}
\end{figure}

Based on the OPT embedding algorithm, ISMark adopts an auto-decoder-like structure instead of previous auto-encoder-like structure, as shown in Fig.\ref{fig2}. In ISMark, the decoder could be any extraction mechanism, for example, decoders pre-trained in previous schemes \cite{zhu2018hidden, liu2019novel}. It's worth noting that a good decoder may result in better performance not only in extraction, but also in embedding. To fully exploit the potential of OPT embedding, we propose a decoder enhancing algorithm to finetune the pre-trained decoder. 

Experiments show that ISMark achieves superior performance in watermarking, exceeding state-of-the-art scheme\cite{fang2022encoded} by 4.64\% (average bit error rate from $4.86\%$ to $0.22\%$) in terms of robustness and by 2.20dB (PSNR from $32.50$dB to $34.70$dB) in terms of invisibility. Compared with baseline\cite{liu2019novel}, ISMark improves error rate from 3.95\% to 0.11\% and improves PSNR from 27.74dB to 37.85dB. In addition, ISMark can embed $4\times$ longer watermarks (bpp from $1.83\times 10^{-3}$ to $7.32\times 10^{-3}$) while keeps high performance (average bit error rate=$2.50\%$ and PSNR=$36.7$dB).

In summary, our technical contributions in this paper are : 
\begin{enumerate}

\item We introduce computational cost in watermarking to lever the watermarking performance, and propose a novel auto-decoder-like image-specified framework. ISMark can effectively exploit the image-specified features to achieve high invisibility and robustness at the same time.

\item We design an optimization-based embedding algorithm to embed robust, invisible and large-capacity watermarks.

\item We propose a decoder enhancing algorithm to future enhance the extraction capability of the decoder.

\item ISMark achieves the state-of-the-art performance on COCO dataset, with an average bit error rate of $0.22\%$ and PSNR of 34.70 dB. 
\end{enumerate}

\section{Related Works}
\label{sec:rela}

\begin{figure*}[t]
    \centering
    \includegraphics[width=\linewidth]{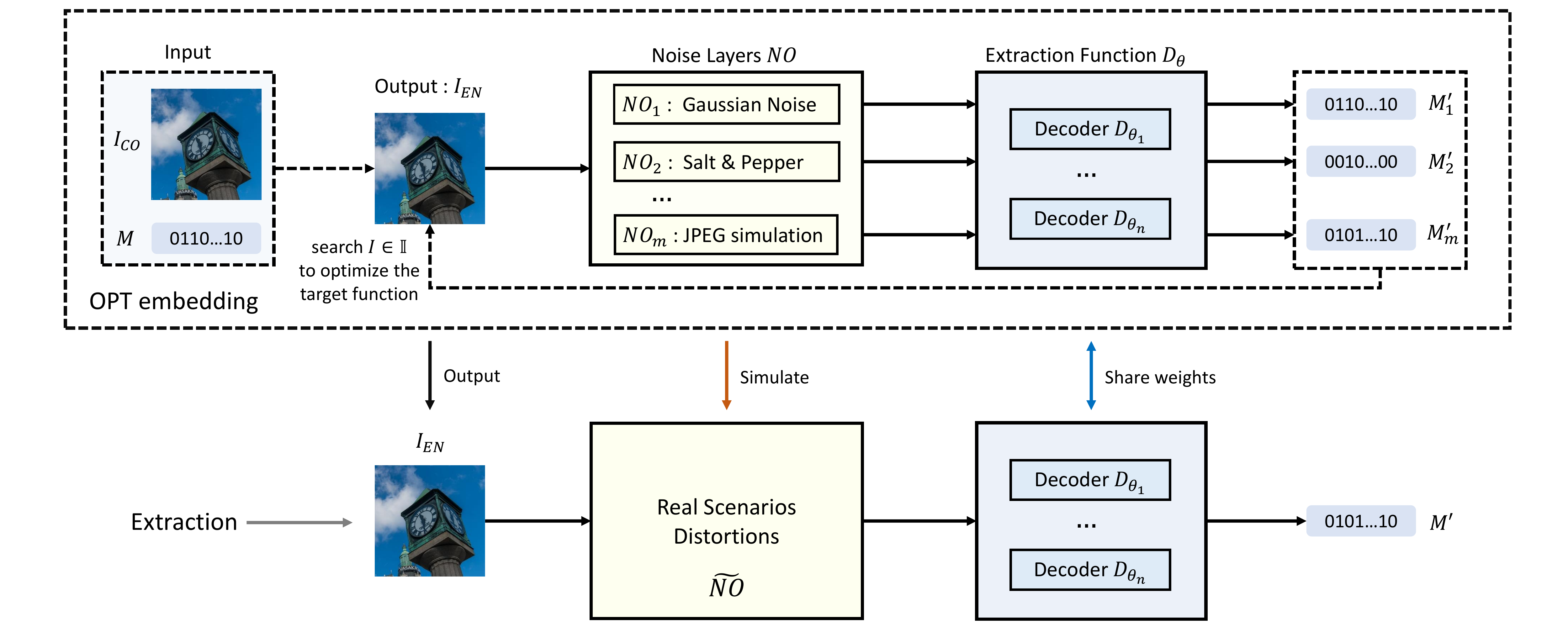}
    
    \caption{An overview of the optimization-based(OPT) embedding algorithm (upper) and the correspondent extraction process (lower). OPT embedding aims to embed watermark $M$ into the cover image $I_{CO}$ and generate the encoded image $I_{EN}$ as output. Given the simulated noise layers $NO$ and decoder $D_\theta$, the encoded image is optimized to minimize the objective function Eq.\ref{equ10}. The optimization is performed on each given cover image and watermark, resulting in full exploitation of image-specified features and high performance.}
    \label{fig3}
\end{figure*}

\subsection{Traditional watermarking schemes}

The watermarking technology is first researched by Ron van Schyndel in 1994\cite{van1994digital}. They propose to embed the watermark in the least significant bit (LSB) of the image pixels, which guarantees the invisibility but cannot survive from distortions. To improve the robustness, many algorithms have been proposed to embed the watermarks in the frequency domain. They focus on the specific transform coefficients of the frequency transform, such as DCT domain\cite{fang2018screen,kang2010geometric}, DFT domain\cite{hamidi2018hybrid} and wavelet domain\cite{bao2005image, bi2007robust,makbol2013robust,wang1998wavelet}. Recently, A symmetry-based watermark synchronization process has been proposed to resist local geometric distortions in spatial domain\cite{ma2021local}. These methods greatly enhance the robustness, but the performance is still limited by the shallow handcrafted features.

\subsection{DNN-based watermarking schemes}

In recent years, with the development of the deep learning algorithms, many DNN-based watermarking schemes have been proposed\cite{zhu2018hidden,ahmadi2020redmark,liu2019novel,zhang2020towards,tancik2020stegastamp}. Zhu et al.\cite{zhu2018hidden} propose a DNN-based auto-encoder-like architecture to jointly train the encoder and decoder with a noise layer, and then many training strategies were proposed to solve existing problems in such DNN-based watermarking schemes. Liu et al.\cite{liu2019novel} proposed a two-stage separable algorithm for robustness against arbitrary noises, and Jia et al.\cite{jia2021mbrs} use real and simulated JPEG compression in different mini-batch to enhance the robustness of JPEG compression. Recently, an encoded feature enhancement scheme is proposed by Fang et al.\cite{fang2022encoded}, which enhances the watermark signal in the Fourier-transform domain and greatly improves the robustness. However, all these schemes are limited by the robustness-invisibility trade-off, which cannot adapt to the scenarios where both high visual quality and robustness are required. It motivates us to go deep to exploit image-specified features to break the robustness-invisibility trade-off.

\subsection{Learning-based Algorithm}
\newpage
With the development of deep learning, many learning-based algorithms have been proposed. Some of them are designed to solve \textit{classification} problems like image classification\cite{he2016deep,liu2021swin} and semantic segmentation\cite{chen2018encoder}, while others are proposed to solve \textit{regression} problems such as denoising\cite{zhang2017beyond} and super-resolution\cite{dong2015image}. These learning-based schemes aim to fit a function by training a neural network on a large-scale dataset, and once the training is complete, the network can predict the result with a single feed-forward pass. Previous learning-based watermark embedding schemes\cite{zhu2018hidden, liu2019novel,jia2021mbrs,fang2022encoded} aim to fit a function to map the cover image domain to the encoded image domain, so they can be categorised as the regression problems.

\subsection{Optimization-based Algorithm}
Unlike learning-based algorithms, optimization-based methods depend on iterative optimization to solve complexity problems and achieve excellent performance. Optimization-based methods are widely adopted in adversarial attack\cite{goodfellow2014explaining, kurakin2018adversarial,chakraborty2021survey}, where the adversary examples can be generated through optimization on the input images to manipulate the predictions of the DNN-based classifiers. And recently, optimization-based algorithms have achieved remarkable performance in image style transfer\cite{gatys2015neural}, novel view synthesis\cite{mildenhall2020nerf}, 3D shape modeling\cite{park2019deepsdf} and image compression\cite{zhao2021universal}, which shows their potential in deep learning. In this paper, we also introduce an optimization-based embedding algorithm for digital watermarking.

\section{Methods}
\label{sec:methods}

Aiming to break the robustness-invisibility trade-off and exploit the features of each specified image, we propose an image-specified watermarking framework (ISMark). The framework adopts an opti-\\mization-based(OPT) embedding algorithm and an enhanced DNN-based decoder.

\subsection{Optimization-based Embedding}
\label{sec:3-1}
Previous watermarking schemes model watermark embedding process as a feed-forward function $E$ to map the cover image $I_{CO}$ and watermark $M$ to the encoded image $I_{EN}$:
\begin{equation}
    E : I_{CO}, M \to I_{EN}
\end{equation}
For learning-based schemes, such functions are modeled by neural networks and then optimized by learning on large-scale datasets. It means that they have to solve a regression problem by learning from the distribution of the datasets, but the feature of each specified image is not fully exploited. To address such limitations, we regard embedding as an optimization process and propose a novel optimization-based (OPT) algorithm.

Given a decoder $D_\theta$ with sparameters $\theta$, a cover image $I_{CO}$, and a watermark $M$, the OPT embedding algorithm searches for the optimal encoded image $I_{EN}$ in all possible images $I\in\mathbb{I}$:
\begin{equation}
    \label{equ2}
    I_{EN} = \arg \min_{I\in \mathbb{I}} L(I; I_{CO}, M)
\end{equation}
Here $\mathbb{I}$ denotes the selection of all possible pixel values, the watermark $M\in\{0,1\}^S$ is represented as a set of bits of length $S$, and $L$ is the objective function. An overview of OPT embedding is depicted in Fig.\ref{fig3}.


\subsubsection{Watermark Extraction Process}
\label{sec:3-1-1}

OPT embedding is based on optimization on the extraction process, so we first define the extraction process of $D_\theta$. In ISMark, $D_\theta$ is a set of $n$ neural networks with different parameters $\theta=\{\theta_1, \theta_2, \dots, \theta_n\}=\{\theta_i\}, i\in \{1,2,\dots,n\}$, each of which can extract a part of the watermark $M_i$. The whole watermark can be written as $M=M_1\ |\ M_2\ |\ \dots\ |\ M_n$, so the total extracting function can be formulated as :
\begin{equation}
    \label{equ5}
    D_{\theta}(I) = D_{\theta_1}(I)\ |\ D_{\theta_2}(I)\ |\ \dots\ |\ D_{\theta_n}(I)
\end{equation}
where signal "$|$" denotes concatenation of bits. 

In practice, an image may be distorted during transmission or processing, resulting in a noised image $I_{NO}$. So the decoder $D_{\theta}$ must extract robustly from $I_{NO}$ to generate $M' = D_{\theta}(I_{NO})$. We can simulate different kinds of distortions by a set of $m$ noise layers $NO=\{NO_1, NO_2, \dots, NO_m\}=\{NO_j\}, j\in\{1, 2, \dots, m\}$, so the noised images can be simulated by $I_{NO_j}=NO_j(I)$. With all above definitions, we can formulate the extraction process in OPT embedding as :
\begin{equation}
    \label{equ6}
    M_j' = D_{\theta}(I_{NO_j}) = D_{\theta}(NO_j(I))
\end{equation}
where $M_j'\in\{0,1\}^S$ denotes watermark extracted from image $I$ with noise layer $NO_j$. We will demonstrate our settings of noise layers in Section \ref{sec:4-1}


\subsubsection{Optimization Objective Function}
\label{sec:3-1-2}
The basic requirements of watermarking algorithms are robustness, invisibility and capacity. In OPT embedding, the capacity is determined by the number of decoder $n$ and the capacity of each decoder $S_i$. The optimization objective is to maximize robustness and invisibility given a watermark capacity $S$. 

The robustness of watermark can be measured by the extraction error under different distortions. As discussed in Section \ref{sec:3-1-1}, we simulate $m$ kinds of distortions with noise layers $NO=\{NO_j\}, j\in\{1,2,\dots,m\}$. For each noise layer $NO_j$, the extraction error is defined as the $L_2$ distance between the watermark $M$ and the extracted watermark $M'_j=D_{\theta}(NO_j(I))$, so the total extraction error can be defined as the average extraction error of these distortions:
\begin{equation}
    \label{equ8}
    L_D = \frac{1}{m}\sum_{j=1}^{m} MSE(M,M'_j)
\end{equation}
The invisibility of watermark can be measured with the $L_2$ distance between the cover image $I_{CO}$ and image $I$ :
\begin{equation}
    \label{equ9}
    L_E = MSE(I_{CO}, I)
\end{equation}
So the optimization objective function is :
\begin{equation}
    \label{equ10}
    L_{\theta} = \lambda_D L_D + \lambda_E  L_E
\end{equation}
where $\lambda_E$ and $\lambda_D$ are the weight factors to balance the trade-off between robustness and invisibility. We can reformulate Eq.\ref{equ2} with $L_\theta$ for the final optimization objective:
\begin{equation}
    \label{equ11}
    I_{EN} = \arg \min_{I\in \mathbb{I}} L_{\theta}(I; I_{CO}, M)
\end{equation}


\subsubsection{Optimization Strategy}

Since the decoder $D_\theta$ and noise layers $NO$ are differential, we can optimize $L_{\theta}(I;I_{CO},M)$ through standard back-propagation algorithm\cite{rumelhart1986learning}. The detailed process is demonstrated in Algorithm \ref{alg:Framwork}. Given cover image $I_{CO}$ and watermark $M$, $I$ is initialised as $I_{CO}$ to speed up the convergence (line 1). The objective function $L_\theta$ is computed by Eq.\ref{equ10} (line 3-8), and then back-propagation algorithm is adopted to compute the gradient and update $I$ (line 9). Following such steps, we update $I$ for $k$ iterations and generate $I_{EN}=I$ as output (line 11). $I$ can be updated by SGD\cite{cherry1998sgd} or more advanced momentum-based algorithms, for example, Adam\cite{kingma2014adam}. For simplification, we only introduce SGD in Algorithm \ref{alg:Framwork}.


\begin{algorithm}[t]
\caption{OPT embedding. The number of iteration $k$, and the optimization step length $l_{OPT}$ are pre-defined hyperparameters. $m$ is the number of the noise layers. The loss function $L_{E}$ and $L_{D}$ and their weight factors $\lambda_E$ and $\lambda_D$ are demonstrated in Section \ref{sec:3-1-2}. }
\label{alg:Framwork}
\begin{algorithmic}[1] 

\REQUIRE ~~\\ 
    The cover image, $I_{CO}$;\\
    The watermark, $M$;
    
\ENSURE ~~\\ 
    The encoded image, $I_{EN}$;
    
\STATE $I=I_{CO}$;

\STATE \textbf{for} $k$ iterations \textbf{do} 

\STATE \ \ \ \ \ \ $M'=\{\}$
\STATE \ \ \ \ \ \ \textbf{for} $j$ in \{1,2,\dots,m\} \textbf{do}
\STATE \ \ \ \ \ \ \ \ \ \ \ \ $M'_j=D_\theta(NO_j(I))$;
\STATE \ \ \ \ \ \ \ \ \ \ \ \ add $M'_j$ to $M'$;
\STATE \ \ \ \ \ \ \textbf{end for}

\STATE \ \ \ \ \ \ $L_{\theta}=\lambda_E L_{E}(I_{CO},I) + \lambda_D L_{D}(M,M')$;
\STATE \ \ \ \ \ \ $I$ =  $I$ - $l_{OPT}\cdot\nabla_{I} L_{\theta}$;
\STATE \textbf{end for}
\STATE $I_{EN}=I$

\RETURN $I_{EN}$; 

\end{algorithmic}
\end{algorithm}

\subsection{Decoder Enhancing}
\label{sec:3-2}

As demonstrated above, the performance of OPT embedding is greatly related to the decoder $D_{\theta}$. So the keypoint is: \textit{how to learn the parameters $\theta$?}

\begin{figure}[t]
    \centering
    \includegraphics[width=0.95\linewidth]{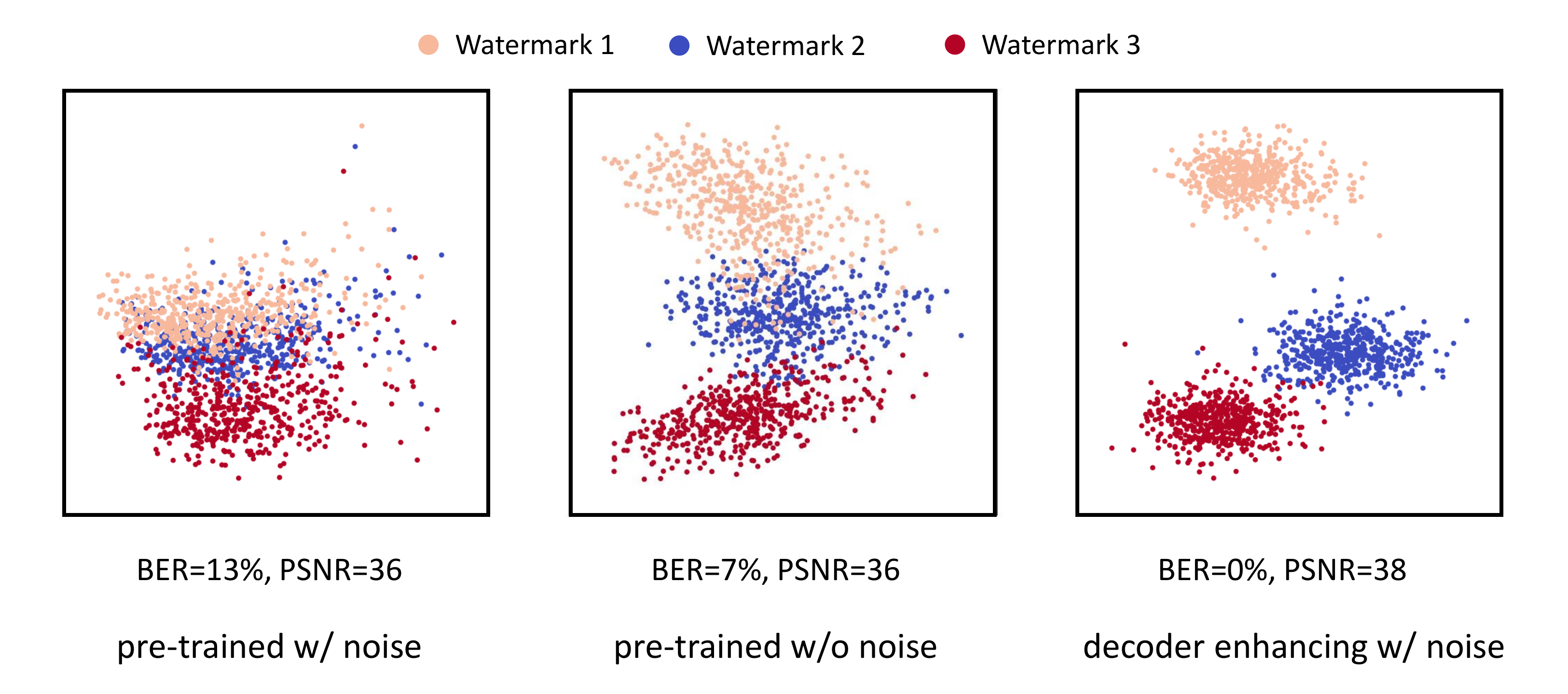}
    
    \caption{Feature visualisation with PCA\cite{pearson1901liii}. We embed three kinds of watermarks (shown in different colors) into 100 images from COCO dataset\cite{lin2014microsoft} through OPT embedding, and visualize the extracted features of different decoders to compare their discriminative capability. We find that the decoder pre-trained with noise layers(left) cannot discriminate different watermarks, which leads to worse extraction accuracy than pre-training without noise layer(middle). However, through decoder enhancing(right) the decoder can well distinguish watermarks and achieves the best performance.}
    \label{fig4}
\end{figure}

A straightforward idea is following existing learning-based schemes such as \cite{zhu2018hidden, liu2019novel} to train $\theta$ in an auto-encoder-like structure. That is, we can directly adopt the pre-trained decoders in \cite{zhu2018hidden, liu2019novel} for OPT embedding. However, there are two problems when using such pre-trained decoders:

\begin{enumerate}
    \item \textit{The extraction capabilities are not powerful enough to fully exploit the potential of OPT embedding}. In previous schemes, the watermark embedding methods are not robust enough so that the embedded watermark features cannot be well preserved after distortions. As a result, the decoders are trained to extract watermark with distorted watermark features, leading to the lack of capabilities to distinguish different watermarks thus causes a lower optimization upper bound. We visualize it by PCA\cite{pearson1901liii}(Fig.\ref{fig4} left) and result shows that the decoder pre-trained with noise layers cannot discriminate different watermarks well (e.g., blue points and brown points in the figure), which results in poor performance in OPT embedding. However, when we pre-train the decoder with complete information by removing the noise layers(Fig.\ref{fig4} middle), the decoder can better discriminate watermarks thus leads to slight performance improvement. But such a decoder is not trained for resist distortions, so it still suffers from the poor robustness.
    
    \item \textit{Mismatch between simulated noise layers and real scenario distortions}. In OPT embedding, we optimize the extraction error against simulated noise layers, but a mismatch occurs when the simulated distortion differs from the real distortion. For example, we simulate JPEG compression with a differential function\cite{shin2017jpeg}, which differs from JPEG compression in real scenarios and leads to performance drop of the decoder. Typically, OPT embedding obtains strong robustness against the simulated JPEG (with bit error rate of 0\%) but almost fail on real JPEG compression (with bit error rate of 31\%).
\end{enumerate}
In order to eliminate such limitations, a decoder enhancing manner is proposed to train a powerful decoder $D_\theta$ for OPT embedding. In the decoder enhancing scheme, OPT embedding is adopted to generate training data with complete watermark information, and the decoder is trained on the noised images distorted by real distortions for strong robustness. As a result, the effect of mismatch is partially removed, and the extraction capability of decoders is well enhanced. As shown in right of Fig.\ref{fig4}, through decoder enhancing, the decoder can fully distinguish different watermarks and the performance is much better than using the pre-trained decoders.

We can denote the distortions in real scenarios as $\widetilde{NO}=\{\widetilde{NO}_j\}, j\in\{1, 2, \dots, \widetilde{m}\}$ , so the extraction process in decoder enhancing can be formulated as :
\begin{equation}
    \label{equ12}
    \widetilde{M'_j} = D_{\theta}(\widetilde{NO}_j(I)),\ \  j\in\{1, 2, \dots, \widetilde{m}\}
\end{equation}
And the loss function is defined as the extraction error :
\begin{equation}
    \label{equ13}
    L = MSE(M,\widetilde{M'_j}),\ \  j\in\{1, 2, \dots, \widetilde{m}\}
\end{equation}

The details of decoder enhancing are shown in Algorithm \ref{alg:IT}. $\theta$ is initialized as a pre-trained decoder $\theta_0$ (line 1), then $\theta_0$ is refined for $K$ iterations. In each training iteration, a batch of cover images $I_{CO}$ and watermarks $M$ are given to generate the encoded images $I_{EN}$ by OPT embedding with $\theta$ (line 3-4). $D_\theta$ is updated for $l$ sub-iterations to minimize loss function $L$ in Eq. \ref{equ13} (line 6-9), in each sub-iteration the real scenario distortion $\widetilde{NO}_j$ is randomly selected from all distortions $\widetilde{NO}$ to generate the noised images.

Once the decoder enhancing manner finishes, the parameters $\theta$ will be fixed. Then for any given images, we can use the same decoder $D_\theta$ to embed and extract watermarks. It means such decoder enhancing algorithm will not increase the computational cost in embedding.

\begin{algorithm}[t]
\caption{Decoder enhancing algorithm. The number of iterations $K$ and the number of sub-iterations $l$ are pre-defined hyperparameters. The loss function $L$ are defined in Eq.\ref{equ13}. }
\label{alg:IT}
\begin{algorithmic}[1] 

\REQUIRE ~~\\ 
    All cover images in the dataset;\\
    An initial parameter $\theta_0$;
    
\ENSURE ~~\\ 
    The enhanced parameter $\theta$;
    
\STATE Initialize $\theta=\theta_0$;

\STATE \textbf{for} $K$ iterations \textbf{do} 

\STATE \ \ \ \ \ \ random select $I_{CO}$ and $M$;
\STATE \ \ \ \ \ \ $I_{EN} = $ OPT embedding $(I_{CO},M;\theta)$;
\STATE \ \ \ \ \ \ \textbf{for} $l$ iterations \textbf{do}
\STATE \ \ \ \ \ \ \ \ \ \ \ \ random select $j\in\{1,2,\dots,\widetilde{m}\}$;
\STATE \ \ \ \ \ \ \ \ \ \ \ \ $\widetilde{M'_j} = D_{\theta}(\widetilde{NO}_j(I))$;
\STATE \ \ \ \ \ \ \ \ \ \ \ \ $L = MSE(M,\widetilde{M'_j})$;
\STATE \ \ \ \ \ \ \ \ \ \ \ \ $\theta = \theta - lr\cdot\nabla_{\theta} L$;
\STATE \ \ \ \ \ \ \textbf{end for}
\STATE \textbf{end for}
\RETURN $\theta$;

\end{algorithmic}
\end{algorithm}

\begin{figure*}[t]
    \centering
    \includegraphics[width=\linewidth]{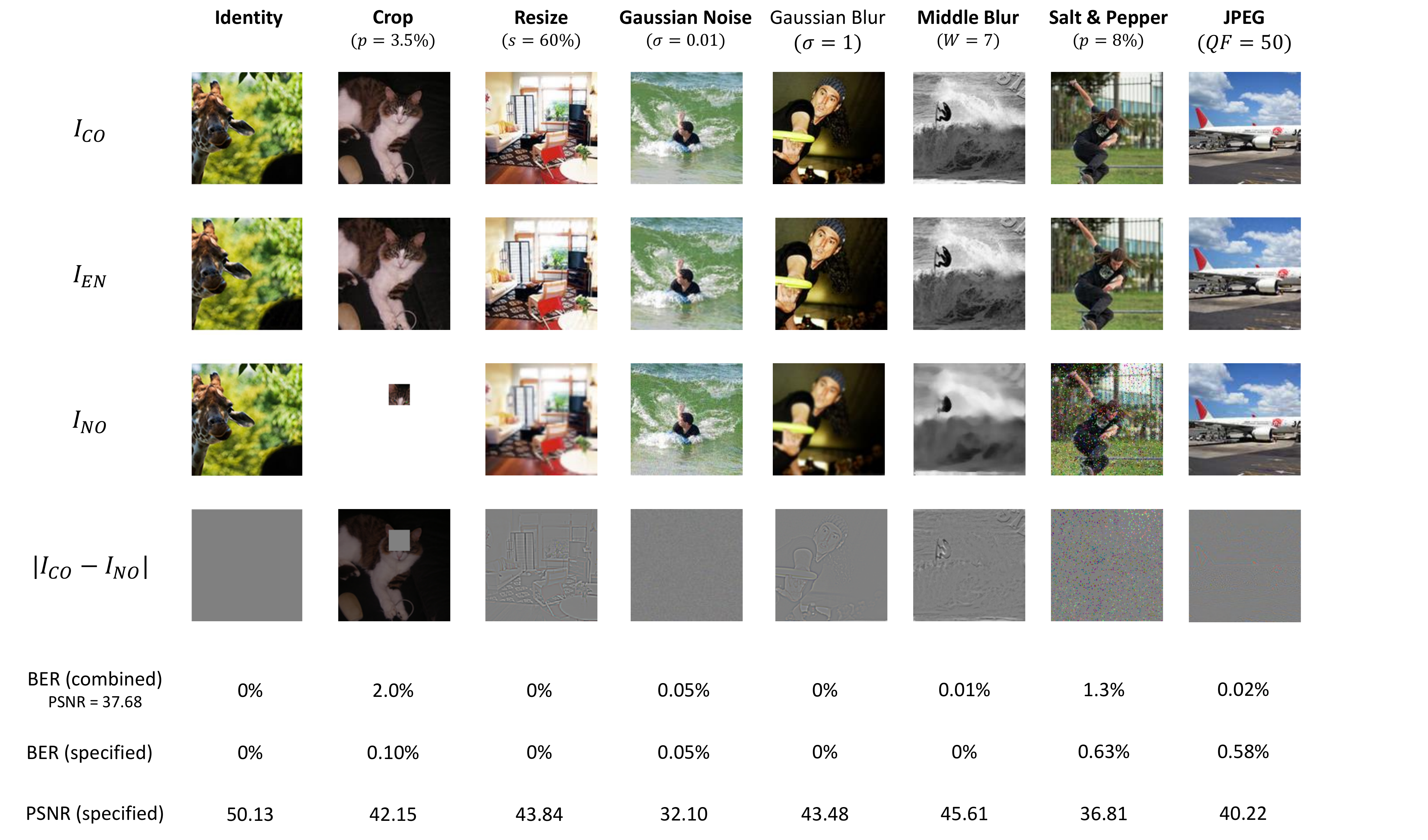}
    
    \caption{The testing results of our IS-HiDDeN model. We show the visual quality and the distortions by 8 images, which are randomly selected from COCO datasets. The cover image $I_{CO}$, the encoded images $I_{EN}$ (generated by the \textbf{combined}-noise model), the noised image $I_{NO}$ and the residual $|I_{CO}-I_{NO}|$ are displayed in the upper of figure. We show the testing results for both \textbf{combined} and \textbf{specified} cases. Bit error rate (BER) and peak signal to noise ratio (PSNR) are displayed in the bottom of figure, and both are the average value measured on the test dataset.}
    \label{fig8}
\end{figure*}

\section{Experiments}
\label{sec:exps}

\begin{table*}
    \caption{Comparison with SOTA. BER(\%) and PSNR are shown in the table. EFE\cite{fang2022encoded} are not tested on crop attack because it is not designed for such robustness. }
    \label{tab1}
	\centering
    \resizebox{0.85\linewidth}{!}{
	\begin{tabular}{c|c|c|c|c c c c|c c c c|c c c}
	    \midrule
		\multirow{2}*{Model} &  \multirow{2}*{bpp ($\times 10^{-3}$)} & \multirow{2}*{PSNR(dB)} &Average & \multicolumn{4}{c}{Crop $p$} & \multicolumn{4}{|c}{Resize $s$} & \multicolumn{3}{|c}{Gaussian Noise $\sigma$}\\
		~ & ~ & ~ & BER & 3.5\% & 16\% & 36\% & 64\% & 0.6 & 0.8 & 1.2 & 1.4 & 0.01 & 0.005 & 0.001 \\
	    \midrule
	    \midrule
		
		HiDDeN\cite{zhu2018hidden} & 1.83 & 27.81  & 6.59 & 7.4 & 2.5 & 2.1 & 2.0 & 1.8 & 2.2 & 1.8 & 1.8 & 2.8 & 2.1 & 1.7\\
		TSDL\cite{liu2019novel}  & 1.83 & 27.74  & 3.95 &  6.2 & 0.3 & 0.2 & 0.1 & 0.2 & 0.1 & 0.1 & 0.1 & 1.3 & 0.3 & 0.1\\
	    \midrule
		IS-HiDDeN  &  1.83 & 37.68  & 0.15 & 2.0 & 0 & 0 & 0 & 0 & 0 & 0 & 0 & 0.05 & 0 & 0.02 \\
		IS-TSDL  & 1.83 & 37.85 & 0.11 & 1.5 & 0 & 0 & 0 & 0 & 0 & 0 & 0 & 0.2 & 0.03 & 0 \\
	    \midrule
	    \midrule
	    
		LGDR\cite{ma2021local}  &3.91 & 32.60 & 12.76 & 33.8 & 22.9 & 12.9 & 9.6 & 6.3 & 4.6 & 3.8 & 3.8 & 7.2 & 7.6 & 7.3\\
		EFE\cite{fang2022encoded} & 3.91 & 32.50  & 4.86 & - & - & - & - & 1.6 & 1.6 & 0.9 & 1.3 & 9.5 & 6.7 & 7.2\\
	    \midrule
		IS-HiDDeN  & 3.91 & 34.60  & 0.18 & 2.9 & 0 & 0 & 0 & 0 & 0 & 0 & 0 & 0.1 & 0 & 0.1 \\
		IS-TSDL  &  3.91 & 34.70 & 0.22 & 2.4 & 0 & 0 & 0 & 0.1 & 0 & 0 & 0 & 0.7 & 0 & 0 \\
	    \midrule
	\end{tabular}
	}
	
	\vspace{10pt}
	
    \resizebox{0.7\linewidth}{!}{
	\begin{tabular}{c|c|c c c|c c c|c c c|c c c}
	    \midrule
		\multirow{2}*{Model} & \multirow{2}*{bpp ($\times 10^{-3}$)} & \multicolumn{3}{c}{Gaussian Blur $\sigma$} & \multicolumn{3}{|c}{Middle Blur $w$}  &  \multicolumn{3}{|c}{Salt\&Pepper $p$} & \multicolumn{3}{|c}{JPEG $Q$}\\
		~ & ~ & 1 & 0.5 & 0.2 & 7 & 5 & 3  & 8\% & 5\% & 2\% & 50 & 70 & 90\\
	    \midrule
	    \midrule
		
		HiDDeN\cite{zhu2018hidden} & 1.83 &  2.6 & 1.9 & 2.1 & 16.0 & 8.7 & 1.9 & 7.5 & 3.0 & 1.7 & 33 & 30 & 15\\
		TSDL\cite{liu2019novel} & 1.83 &  0.5 & 0.1 & 0.1 & 13.6 & 3.7 & 0.2 & 4.7 & 0.9 & 0.1 & 28 & 22 & 8 \\
	    \midrule
		IS-HiDDeN &  1.83 & 0 & 0 & 0 & 0.01 & 0 & 0 & 1.3 & 0.06 & 0 & 0.02 & 0.01 & 0.01\\
		IS-TSDL  &  1.83 & 0 & 0 & 0 & 0 & 0 & 0 & 0.6 & 0.08 & 0 & 0.04 & 0.03 & 0.03\\ 
	    \midrule
	    \midrule

		LGDR\cite{ma2021local} & 3.91 &16.2 & 9.8 & 4.6 & 47.3 & 30.3 & 12.7 & 7.9 & 10.5 & 7.1 & 10.0 & 9.6 & 7.8 \\
		EFE\cite{fang2022encoded} &3.91 & 7.9 & 9.6 & 1.2 & 6.0 & 5.5 & 4.7 & 3.1 & 3.0 & 2.7 & 8.5 & 6.3 & 5.0 \\
	    \midrule
		IS-HiDDeN & 3.91 &0 & 0 & 0 & 0 & 0 & 0 &  1.2 &  0.1 &  0 &  0 &  0 &  0 \\
		IS-TSDL  & 3.91 &0 & 0 & 0 & 0 & 0 & 0 & 1.3 & 0.3 & 0 & 0.3 & 0.1 &  0.1\\
	    \midrule
		
	\end{tabular}
	}
\end{table*}






\subsection{Implementation Details}
\label{sec:4-1}

In OPT embedding, we optimize $I_{EN}$ for $k=200$ iterations by default. We simulate 7 kinds of normal distortions as the noise layers $NO$, including \textit{crop, resize, Gaussian noise, Gaussian blur, middle blur, salt and pepper, and JPEG compression}. Details of the definition and parameters of these noises are demonstrated in the supplementary material. An additive \textit{identity} layer is also adopted, which sets no distortion on the encoded image. For the weight factors of the objective function Eq.\ref{equ10}, we choose $\lambda_E=1$ and $\lambda_D=2$. We use Adam algorithm\cite{kingma2014adam} with a learning rate of $10^{-3}$ to update $I_{EN}$ by default.

In decoder enhancing, the decoder is trained on 10,000 images from the ImageNet dataset\cite{deng2009imagenet}. Watermarks are sampled randomly at each bit. $n$ decoders are trained separably to save memory. We use Adam algorithm with the same learning rate at in pre-training ($10^{-3}$ for IS-HiDDeN and $10^{-4}$ for IS-TSDL, as introduced in Section \ref{sec:4-3}) to optimize the decoder. The batch size is set to 6. The decoder is optimized for $K=10000$ iterations, each with $l=80$ sub-iterations, which takes about 15 days on one single NVIDIA RTX 2080Ti GPU. More training details can be found in the supplement material.

\subsection{Evaluation}
\label{sec:4-2}

ISMark is evaluated on 500 images from COCO dataset\cite{lin2014microsoft}. We use the \textbf{b}it \textbf{e}rror \textbf{r}ate (BER) of the extracted watermarks to measure the robustness, use PSNR to measure the invisibility, and use \textbf{b}it \textbf{p}er \textbf{p}ixel (bpp) to measure the watermark capacity. The executing time is also measured for the OPT embedding, which is reported in Section \ref{sec:4-6-4}.

\subsection{Baseline}
\label{sec:4-3}

The baseline for comparison are HiDDeN\cite{zhu2018hidden}, TSDL\cite{liu2019novel}, LGDR\cite{ma2021local} and EFE\cite{fang2022encoded}. LGDR is the SOTA traditional watermarking scheme, and EFE is the SOTA learning-based scheme. HiDDeN and TSDL are two learning-based watermarking schemes and we train their decoders without noise layers for 300 epochs as the initial paramterers $\theta_0$ of ISMark. According to the backbone of the decoder, we call models trained by our scheme as IS-HiDDeN and IS-TSDL correspondingly.

To verify the capability of ISMark against non-differential distortions, we also compare with ASL\cite{zhang2020towards}, StegaStamp\cite{tancik2020stegastamp} and MBRS\cite{jia2021mbrs} that achieve the state-of-the-art performance in terms of robustness of JPEG compression. We also adopt MBRS as backbone to train a IS-MBRS for comparison.

\subsection{Quantitative Results}
\label{sec:4-4}

First we perform the quantitative results to evaluate the performance of our method. As demonstrated in Section \ref{sec:4-1}, we use 7 kinds of distortions and an additive identity layer for both OPT embedding and decoder enhancing. Similar to \cite{liu2019novel}, We train 8 \textbf{specified} models each adopting one noise layer to obtain specified robustness, and a \textbf{combined} model to adopt all 8 layers for robustness against all distortions.  Fig.\ref{fig8} shows some examples of high-intensity noises with the robustness and invisibility of the IS-HiDDeN model. 

As shown in Fig.\ref{fig8}, we have achieved both high invisibility and strong robustness for all distortions. In particular, our model achieves BER$<0.7\%$ with a average PSNR of $41.79$dB for all specified noises, and achieves BER$\le 2\%$ with PSNR$=37.68$dB for combined noises, which shows the high performance of our method. Beneficial from the decoder enhancing algorithm, our model gets strong robustness against non-differential JPEG Compression with BER=$0.02\%$ in combined case. And we also note that the robustness is little weaker for the crop attack and salt \& pepper noise, with BER = $2.0\%$ and $1.3\%$ correspondingly. It is partly because these noises are independent to the pixel value of image thus the gradient computed on the distorted pixels are zero, which makes it difficult to optimize the extraction error. Still they are much stronger than the previous methods, as is shown in the following section.

\begin{figure}[h]
    \centering
    \includegraphics[width=0.8\linewidth]{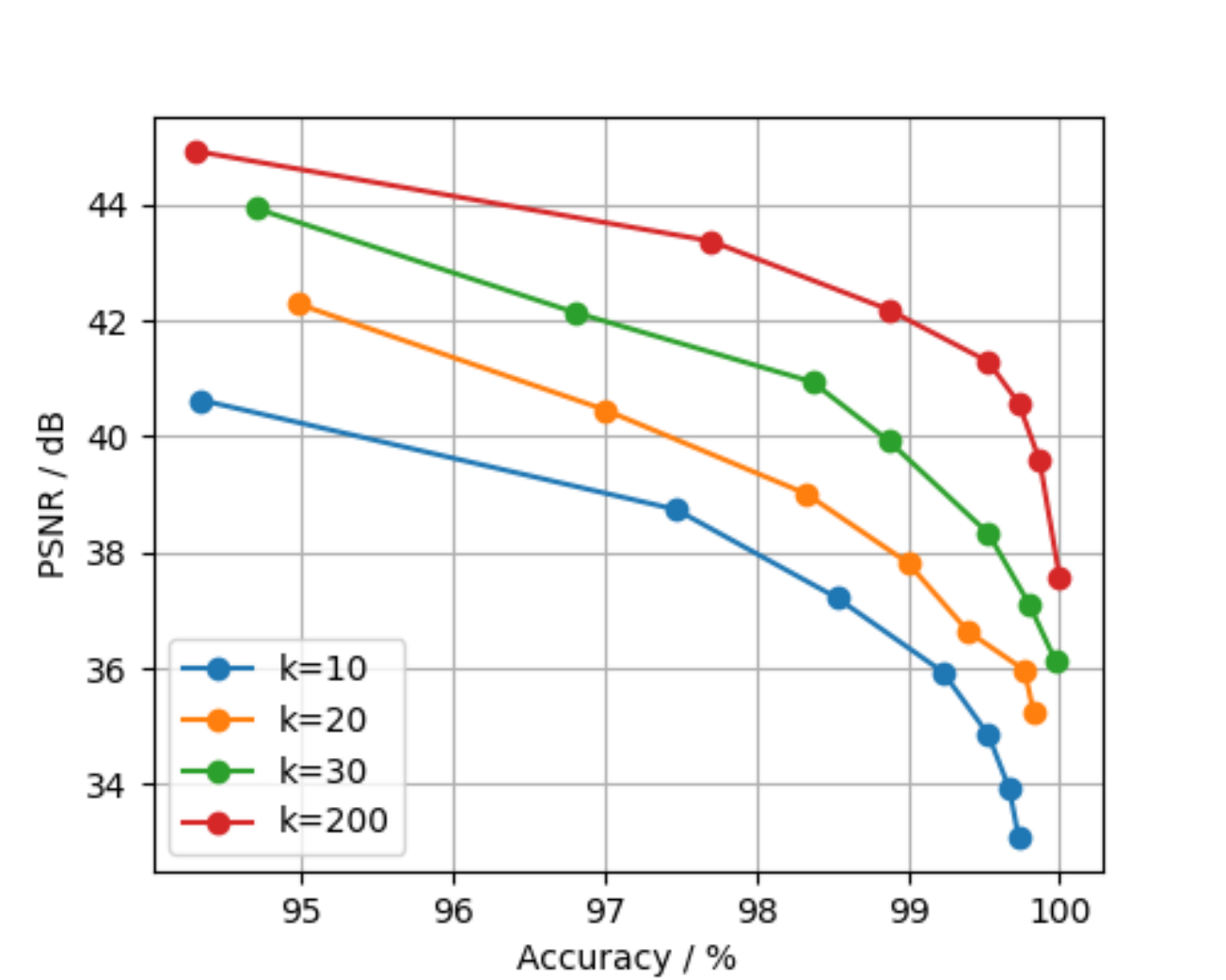}
    \caption{Levering watermarking performance with computational cost.}
    \label{fig10}
\end{figure}

We further perform an experiment to lever the watermarking performance by increasing computational cost. We test IS-TSDL under Gaussian noise ($\sigma=0.01$) with optimization step $k\in\{10, 20, 30, 200\}$, and for each $k$ we change the optimization step length $l_{OPT}$ to plot the PSNR-Accuracy curve. As shown in Fig.\ref{fig10}, with the increment of $k$, the watermarking performance is well enhanced. It means our method can achieve better performance by simply increasing the optimization step $k$. 

\subsection{Comparison with SOTA}
\label{sec:4-5}

In this section, we perform a comparison experiments with the SOTA methods. We first compare ISMark with HiDDeN\cite{zhu2018hidden}, TSDL \cite{liu2019novel}, LGDR\cite{ma2021local} and EFE\cite{fang2022encoded} under combined noises setting. When $bpp=1.83\times10^{-3}$ the optimization step is set to $k=200$, while for $bpp=3.91\times10^{-3}$ it is set to $k=300$ since it is more difficult to embed larger watermarks.

As is shown in Table.\ref{tab1}, our models exceeds SOTA in both robustness and invisibility by a large margin. We show an average improvement of 4.64\% in terms of BER (from 4.86\% of EFE to 0.22\% of IS-TSDL) while the PSNR increases by approximately 2.20 dB (from 32.50 dB of EFE to 34.70dB of IS-TSDL). The improvement mainly comes from the fact that our method can exploit the image-specified features for watermark embedding, while baselines only utilize the statistic clues.

\begin{table}
    \centering
    \caption{Comparison with SOTA under JPEG compression with quality factor $Q=50$.}
    \label{tab2}
    \resizebox{0.8\linewidth}{!}{
	\begin{tabular}{c|c|c|c}
	    \midrule
		Model &  bpp ($\times 10^{-3}$) & PSNR(dB) & BER(\%)\\
	    \midrule
	    ASL\cite{zhang2020towards} & 1.83 & 27.84 & 9.27 \\
	    StegaStamp\cite{tancik2020stegastamp}& 0.63 & 32.36 & 0.24 \\
	    MBRS\cite{jia2021mbrs} & 3.91 & 36.50 & 0.0092 \\
	    \midrule
		IS-HiDDeN  & 1.83 & 37.68 & 0.02\\
		IS-TSDL & 1.83 & \textbf{37.86} & 0.04\\
		IS-MBRS & 3.91 & 37.43 & \textbf{0.0031}\\
	    \midrule
	\end{tabular}
    }
    \label{sm-tab3}
\end{table}

We also compare with ASL\cite{zhang2020towards}, StegaStamp\cite{tancik2020stegastamp} and MBRS\cite{jia2021mbrs} under JPEG compression specified setting. As is shown in Table.\ref{tab2}, our IS-HiDDeN and IS-TSDL can achieve better performance than ASL and StegaStamp but is a little weaker than MBRS in terms of robustness. However, our IS-MBRS that utilize the MBRS as decoder backbone even outperforms MBRS by 0.0061\% in terms of BER and 0.97 dB in terms of PSNR, which shows the superiority of ISMark.

\subsection{Ablation Study}
In ISMark there are lots of hyper-parameters, so we perform an ablation study to show their effects. We study four factors : decoder enhancing algorithm, the sub-optimization iteration $l$ in decoder enhancing algorithm, the watermark capacity $S$, and the computation cost. 

\begin{figure}
    \centering
    \includegraphics[width=\linewidth]{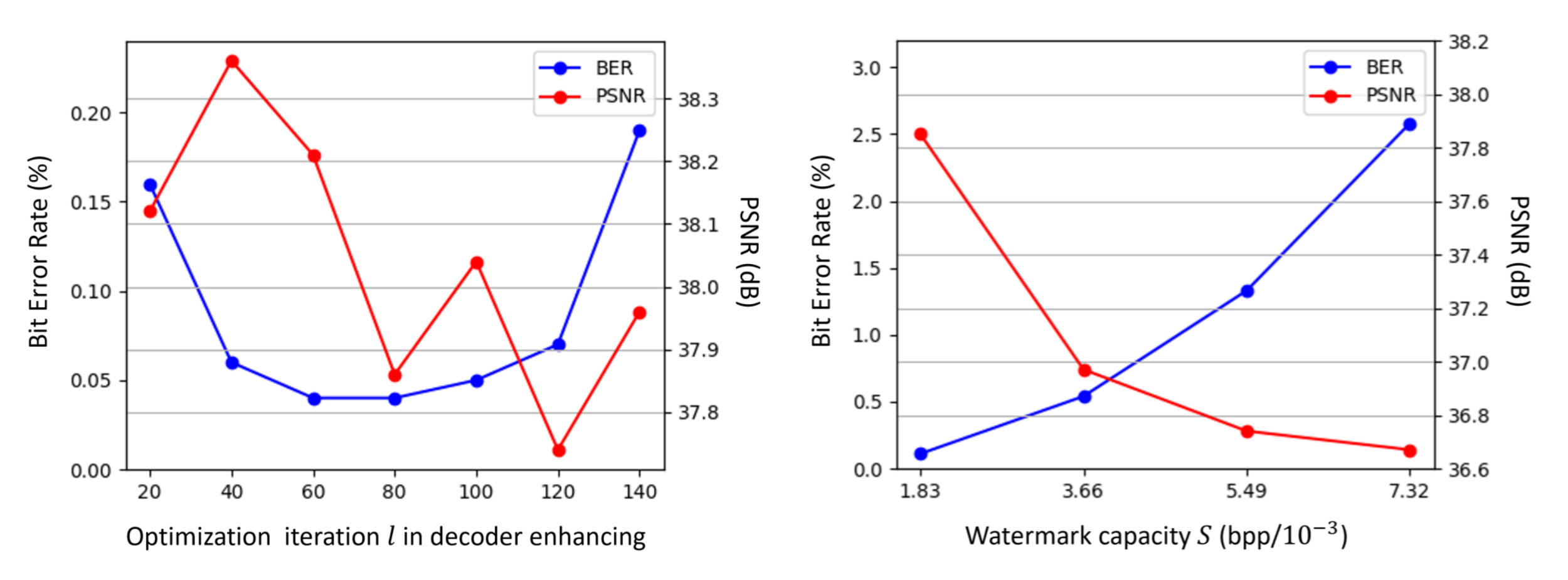}
    \caption{Ablation study on the optimization iteration $l$ in decoder enhancing (left) and the watermark capacity $S$ (right).}
    \label{fig7}
\end{figure}

\subsubsection{Decoder enhancing algorithm}
In decoder enhancing, we optimize the decoder initialized with pre-trained parameters $\theta_0$. To see the effect, we first remove the decoder enhancing and test the performance using the pre-trained $\theta_0$ in OPT embedding. As is shown in Table.\ref{tab3}, ISMark achieves stronger robustness with decoder enhancing, which illustrates the necessity to enhance the pre-trained parameters $\theta_0$. We also train the decoder from scratch, and results show that a randomly initialed decoder can also achieve high robustness and invisibility, while training with pre-trained parameters $\theta_0$ achieves the best performance.

\subsubsection{Sub-optimization iteration $l$ in decoder enhancing}
\label{sec:4-6-2}

In decoder enhancing, the decoder is optimized for $l$ sub-iterations for each batch. We change $l$ from 20 to 140 with an interval of 20, and plot the PSNR and average BER in the left of Fig.\ref{fig7}. We find that with the increase of $l$, the robustness first increases and then decreases, and the best performance is reached around $l=80$. And there is no obvious change for PSNR, which varies from 37.74dB to 38.36dB. This means that the change in iteration number has no obvious effect on invisibility. 

\subsubsection{Watermark capacity $S$}
\label{sec:4-6-3}

As demonstrated in Section \ref{sec:3-1-1}, we can use a set of decoders to embed multiple watermarks, leading to different capacity $S$. We test OPT embedding with capacity $bpp=1.83\times 10^{-3}, 3.66\times 10^{-3}, 5.49\times 10^{-3}, 7.32\times 10^{-3}$ bits, and the results are shown in right of Fig.\ref{fig7}. We find that our ISMark achieves high robustness and invisibility even with a large capacity of $bpp=7.32\times 10^{-3}$ bits. Both metrics decrease as $S$ increases, because it is difficult to guarantee the performance of large-capacity watermarks.

\begin{table}
    \centering
    \caption{Ablation study of decoder enhancing algorithm. For each item we show PSNR(dB)/Average BER in the table.}
    \label{tab3}
    \resizebox{0.9\linewidth}{!}{
	\begin{tabular}{c|c c}
    \midrule
	model&IS-HiDDeN&IS-TSDL\\
	\midrule
	
	 w/o decoder enhancing & \textbf{37.73} / 11.76\% & 36.10 / 10.26\% \\
	 w/o pre-trained  & 37.68 / 0.47\% & \textbf{38.12} / 0.46\% \\
	 
	\midrule
	 baseline & \ 37.68 / \textbf{0.15}\% & 37.85 / \textbf{0.11}\% \\
    \midrule
	\end{tabular}
	}
\end{table}

\begin{table}
    \centering
    \caption{Cost time with different optimization steps $k$ in OPT embeding for IS-TSDL.}
    \label{tab4}
    \resizebox{0.8\linewidth}{!}{
	\begin{tabular}{c|c c c c}
	    \midrule
	k & 50 & 100 & 150 & 200 \\
	\midrule
	embedding time(s) & 2.9 & 5.2 & 7.9 & 10.6 \\
	extracting time(s) & 0.005 & 0.005 & 0.005 & 0.005   \\
    \midrule
	\end{tabular}
	}
\end{table}

\subsubsection{Computation cost}
\label{sec:4-6-4}

The computation cost (measured by runtime) can be controlled by changing the optimization steps $k$ in OPT embedding. We test the embedding time and extracting time for different $k$ in Table.\ref{tab4}. We can find the embedding time is about to increase lineally with $k$, while the extracting time is not influenced.

\section{Limitations}
While ISMark enjoys the high robustness under various digital editing methods like crop, resize and JPEG compression, how to obtain robustness against physical attacks (e.g, print or photograph) is still a problem. Many recent works\cite{liu2019novel,fang2022encoded} have focused on such robustness for learning-based watermarking schemes, and we will learn from these methods to further improve ISMark on physical attack robustness.



\section{Conclusion}
In this paper, we introduce computational cost in watermarking to lever the watermarking performance, and propose a novel auto-decoder-like image-specified watermarking framework. We regard watermark embedding as a optimization process, and propose an optimization-based embedding algorithm to fully exploit the image-specified features. With the proposed decoder enhancing algorithm, our method achieves the state-of-the-art performance in terms of robustness, invisibility and watermark capacity. We believe that ISMark can provide a new perspective for practical digital watermarking.

\bibliographystyle{ACM-Reference-Format}
\bibliography{paper}

\end{document}